\documentclass[9pt,twocolumn,twoside]{pnas-new}

\templatetype{pnasresearcharticle} 

\newcommand{\vrms}{v_\text{rms}}
\newcommand{\fcrit}{f_\text{c}}
\newcommand{\Rcrit}{R_\text{c}}
\newcommand{\vp}{v_\text{p}}
\newcommand{\tmix}{t_\text{mix}}
\newcommand{\tfix}{t_\text{fix}}

\renewcommand\b[1]{{\bf  #1}}
\renewcommand\vec[1]{\boldsymbol{#1}}

\newcommand\del{\nabla}

\newcommand\dd{\mathrm{d}}

\title{Turbulent mixing controls fixation of growing antagonistic populations}

\author[a]{Jonathan Bauermann}
\author[b]{Roberto Benzi}
\author[a]{David R.~Nelson}
\author[a,c]{Suraj Shankar}
\author[d]{Federico Toschi}

\affil[a]{Department of Physics, Harvard University, Cambridge, MA 02138, United States of America}
\affil[b]{Department of Physics and INFN, University of Rome Tor Vergata, I-00133 Rome, Italy}
\affil[c]{Department of Physics, University of Michigan, Ann Arbor, MI 48109, USA}
\affil[d]{Department of Applied Physics, Department of Mathematics and Computer Science, Eindhoven University of Technology, 5600 MB Eindhoven, The Netherlands}
\affil[e]{CNR-IAC, I-00185 Rome, Italy}

\leadauthor{Bauermann}

\significancestatement{
From microbes to mammals, living organisms experience conflict and predation, causing their populations to segregate strongly with antagonistic interactions. Similar interactions govern patterns in reacting chemical mixtures, such as in chiral synthesis. But how does this spatial organization occur in a dynamic environment like turbulent flows? Through numerical simulations and analytical calculations, we demonstrate novel consequences of turbulent mixing on mixtures of proliferating organisms that antagonize each other. Survival or extinction of a population proceeds via a biological nucleation process whose threshold is dramatically modified by fluid dynamics. Altogether, our work highlights the impact of turbulence on spatial population genetics with nonreciprocal interactions, with implications for marine microbial ecology, prebiotic life scenarios, and nonequilibrium chemical pattern formation.
}

\authordeclaration{The authors declare no competing interest.}
\correspondingauthor{\textsuperscript{2}To whom correspondence should be addressed. E-mail: drnelson@fas.harvard.edu}

\keywords{spatial population dynamics $|$ antagonism $|$ nonreciprocal interactions $|$ turbulent mixing}

\begin{abstract}
Unlike coffee and cream that homogenize when stirred, growing micro-organisms (e.g., bacteria, baker's yeast) can actively kill each other and avoid mixing. How do such antagonistic interactions impact the growth and survival of competing strains, while being spatially advected by turbulent flows?
By using numerical simulations of a continuum model, we study the dynamics of two antagonistic strains that are dispersed by incompressible turbulent flows in two spatial dimensions. A key parameter is the ratio of the fluid transport time to that of biological reproduction, which determines the winning strain that ultimately takes over the whole population from an initial heterogeneous state.
By quantifying the probability and mean time for fixation along with the spatial structure of concentration fluctuations, we demonstrate how turbulence raises the threshold for biological nucleation and antagonism suppresses flow-induced mixing by depleting the population at interfaces. Our work highlights the unusual biological consequences of the interplay of turbulent fluid flows with antagonistic population dynamics, with potential implications for marine microbial ecology and origins of biological chirality. 
\end{abstract}

\dates{This manuscript was compiled on \today}
\doi{\url{www.pnas.org/cgi/doi/10.1073/pnas.XXXXXXXXXX}}

\begin{document}

\maketitle
\thispagestyle{firststyle}
\ifthenelse{\boolean{shortarticle}}{\ifthenelse{\boolean{singlecolumn}}{\abscontentformatted}{\abscontent}}{}


\dropcap{L}ife in the ocean is responsible for sustaining global ecological and geophysical processes, such as the carbon cycle, of which marine phytoplankton are the primary drivers of this energy and mass flux \cite{siegel2023quantifying}. The growth of these microorganisms occurs in a complex and dynamic fluid environment, naturally raising the question of how turbulent flows impact their spatial distribution \cite{benzi_spatial_2022}. That turbulent flows near the ocean surface can generate patchy spatial patterns \cite{martin2003phytoplankton}, create transient hydrodynamical niches \cite{karolyi2000chaotic,d2010fluid}, and enable coexistence of diverse phytoplankton populations has been recognised across a wide range of spatial ($1-100~$km) and temporal ($\sim$ days-months) scales \cite{karolyi2000chaotic,barton2010patterns,d2010fluid,mcgillicuddy2016mechanisms,mahadevan2016impact,villa2020ocean}. But how these hydrodynamic processes combine with biological interactions between growing communities of distinct microbial strains remains a challenging open question.

Previous studies have investigated the role of both dynamic and static fluid flows on growing, competing populations \cite{perlekar_population_2010,pigolotti_population_2012,pigolotti_growth_2013,benzi_spatial_2022} with a focus on compressible effects that can strongly suppress the carrying capacity \cite{perlekar_population_2010} and modify the effective selective advantage of the population \cite{plummer_fixation_2019,plummer2023oceanic}. While compressible effects are important on the sub-mesoscale ($\sim 1-10~$km spatial scale) due to strong horizontal flow divergence \cite{plummer2023oceanic,martin2003phytoplankton}, on larger scales $\geq 10~$km, flows are approximately two dimensional and incompressible. Furthermore, phytoplankta often exhibit asymmetric and antagonistic interactions, e.g., between bacteria and toxic algae or phages \cite{doucette1995interactions,long2001antagonistic,martiny2014antagonistic}, that cause competing strains to actively attack, poison or kill one another. In spatial settings, such antagonism leads to a novel kind of active matter \cite{hallatschek2023proliferating} with segregated domains and a genetic line tension that prevents intermingling of the different strains \cite{rouhani1987speciation,lavrentovich2019nucleation}. As a result, a biological nucleation process, rather than simple Darwinian selective advantage, governs the fixation or extinction of either strain, as has been recently demonstrated for antagonistic strains of \textit{Saccharomyces cerevisiae} (baker's yeast) competing on a Petri dish \cite{giometto_antagonism_2021}.

Here we combine fluid dynamics with spatial population genetics of a simple reaction-diffusion model to study the role of turbulent mixing on fixation of antagonistically interacting organisms. 
The combination of both processes yields an unusual setting of constantly stirring a genetic mixture that tries to actively demix by itself, influenced by a genetic line tension \cite{giometto_antagonism_2021}. This transient dynamics terminates when one of the strains takes over the entire population and achieves genetic fixation. Remarkably, which strain finally wins depends strongly on the strength of turbulent mixing which shears the mixture; creating more biological interfaces with antagonism depletes the population. We develop an analytical model for the flow-modified nucleation process, and quantitatively test its predictions with our numerical results. Finally, we compute the spatial spectrum of concentration fluctuations and compare them with other phase separating systems. We conclude by suggesting potential experimental tests and highlighting the implications of our results for biological and chemical systems.   

\section*{Reaction-diffusion model for turbulent antagonism}
Following previous studies we adopt a continuum description that neglects demographic noise \cite{crow2017introduction} and models the population dynamics in two dimensions (2D) as \cite{benzi_spatial_2022}
\begin{align}
    (\partial_t +\mathbf{v}\cdot\vec{\del})c_A=D \nabla ^2 c_A+ \mu c_A (1-c_A- c_B+\alpha c_B) \;, \label{eq:dcAdt}\\
    (\partial_t +\mathbf{v}\cdot\vec{\del})c_B=D \nabla ^2 c_B+ \mu c_B (1-c_A- c_B+\beta c_A)\label{eq:dcBdt} \;, 
\end{align}
where $c_A(\b{x},t)$, $c_B(\b{x},t)$ are the local concentrations of the two strains ($A$, $B$) advected by a 2D incompressible turbulent flow field $\b{v}(\b{x},t)$ along with an effective diffusivity $D$. The initial growth rate of both strains is assumed to have a common value $\mu$ for simplicity, and $\alpha,\beta$ parameterize the interactions between strains. We will focus on the case where $\alpha<0$ and $\beta<0$ corresponding to antagonism \cite{lavrentovich2019nucleation}. Note that these interactions are non-reciprocal for all asymmetric settings $\alpha \neq \beta$.

The two-dimensional flow field $\mathbf{v}$ is obtained by considering the evolution of two-dimensional slices of three-dimensional, homogeneous and isotropic turbulent flow field generated from direct numerical simulations of the Navier-Stokes equation and retaining only the incompressible part (see S.I. for details).
For simplicity, we ignore any feedback from the concentration fields on the hydrodynamic flows and consider equal diffusion constants for both strains, leaving the study of these interesting effects for the future. 

Stirred turbulent flows of a typical strength $\vrms$ (the root-mean-square velocity) on the scale of the system size $L$ are controlled by the Reynolds number $\text{Re}=\vrms L/\nu$ where $\nu$ is the kinematic viscosity of the fluid. We restrict ourselves to a Reynolds number of $\text{Re}\approx100$ in all the calculations in this work for reasons of computational efficiency.
Upon comparing the large eddy turnover time $\tau_\text{eddy}=L/\vrms$ \cite{frisch1995turbulence} with the time scale of reproduction ($1/\mu$), we obtain a key dimensionless number, the Damk{\"o}hler number \cite{neufeld2009chemical}
\begin{equation}
    \text{Da}=\dfrac{\mu L}{\vrms} \;, \label{eq:damkoehler}
\end{equation}
which controls the relative importance of hydrodynamic transport to biological growth. The strength of diffusion in population dynamics can be compared to fluid transport either via viscous diffusion (in terms of a Schmidt number $\text{Sc}=\nu/D$) or advection (in terms of a P{\'e}clet number, ${\rm Pe}=\vrms L/D={\rm Re}\;{\rm Sc}$). In this study, we vary the Damk{\"o}hler number (${\rm Da}$) by varying the hydrodynamic time scale $\tau_{\rm eddy}$ and thus, $\vrms$, while keeping the Reynolds number (${\rm Re}$) and key spatial scales constant, see S.I.~Section~1 for details.

To gain analytical insight into the population dynamics, it is useful to rewrite the model (Eqs.~\ref{eq:dcAdt},~\ref{eq:dcBdt}) in terms of the total concentration of both 
strains $c_T=c_A+c_B$ and the relative fraction of the $A$ strain $f=c_A/c_T$ \cite{benzi_spatial_2022}, which gives
\begin{align}
    (\partial_t+\mathbf{v}\cdot\vec{\del})c_T&=D \nabla^2c_T+\mu c_T[1-c_T-2\sigma\;c_Tf(1-f)] \;,  \label{eq:dcTdt}\\
(\partial_t+\mathbf{v}\cdot\vec{\del})f&=D \nabla^2f+\dfrac{2D}{c_T}\vec{\nabla} c_T\cdot\vec{\nabla}f \nonumber \\
&\quad+\mu c_Tf(1-f)\left[\dfrac{\delta}{2}+\sigma(2f-1)\right]\;. \label{eq:dfdt}
\end{align}
The selective advantage $\delta=\alpha-\beta$ quantifies the growth rate difference between the strains, i.e., for $\delta \gtrless 0$ strain $A$ has a reproductive advantage/disadvantage over the strain $B$, while the nature of the interactions is quantified by $\sigma= -(\alpha+\beta)/2$ ($\sigma<0$: mutualistic, $\sigma=0$: neutral, $\sigma>0$: antagonistic). Here we focus on the antagonistic case and choose $\sigma>0$, $\delta>0$.

One advantage of this parameterization is that for weak enough antagonism, i.e., $\sigma\ll1$, the total concentration is approximately  $c_T\approx 1$ in incompressible flows ($\vec{\del}\cdot\b{v}=0$) \cite{benzi_spatial_2022}. In this limit, \eqref{eq:dfdt} reduces to an advected version of the dynamics of a non-conserved order parameter, so-called Model A dynamics in the classification of Hohenberg and Halperin \cite{hohenberg_theory_1977}. Unlike the more common setting of a conserved order parameter coupled to fluid flow (Model H) \cite{hohenberg_theory_1977} or driven by turbulence (advected Model B)\cite{ruiz1981turbulence,berti2005turbulence,perlekar_spinodal_2014}, e.g., stirring oil and water, the turbulent mixing of a non-conserved order parameter is unusual in condensed matter, as Model A typically describes the dynamics of magnets, which usually lose their magnetism before becoming liquid.

\begin{figure}[t]
\centering
\includegraphics[width=\linewidth]{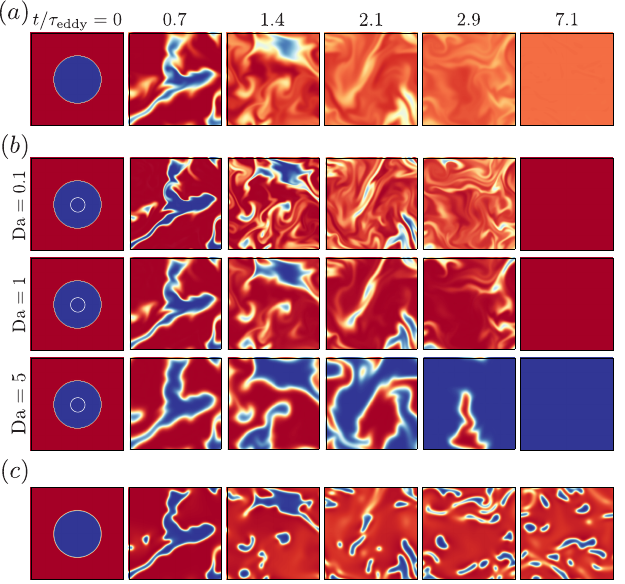}
\caption{Snapshots of the concentration field of a passive scalar in (a), the field of fraction $f$ for different Damk{\"o}hler numbers in (b), and the concentration field of a binary phase separating mixture in (c) under turbulent mixing. All systems are stirred with the same turbulent flow field and shown at the same time points measured in units of the large eddy turnover time ($\tau_{\rm eddy}$). Additionally, we show the critical nucleation radius (white circle) in (b) in the absence of flow. Due to the identical initialization and flow fields, the structures formed in the early times (second and third column) look similar between all three processes. 
In all systems, we use a circular initialization that covers $20\%$ of the system size (blue circle/first column).
In (a) and (c), we show the concentration fields for a freely diffusing and advected passive scalar (``milk'') or a scalar following model B with advection (``oil'').}
\label{fig:examples_mixing}
\end{figure}

In Fig.~\ref{fig:examples_mixing}, we compare the turbulent mixing of a passive scalar (e.g., ``milk in coffee'', Fig.~\ref{fig:examples_mixing}a and Mov.~S1), two antagonistic strains with $\delta=0.1$ (blue has a selective advantage), $\sigma = 0.25$ and varying Damk{\"o}hler number (Fig.~\ref{fig:examples_mixing}b and Mov.~S1), and a phase separating conserved order parameter (e.g., ``oil in water'', Fig.~\ref{fig:examples_mixing}c and Mov.~S1). In all cases, the advecting velocity field and initialization is the same, and it satisfies periodic boundary conditions in the $L\times L$ box. The dynamic patterns obtained are shown at identical time points (in units of $\tau_{\rm eddy}$). As a result, the early time dynamics (second and third column, Fig.~\ref{fig:examples_mixing}) look similar in all five cases.

However, after the initial transient, the freely diffusing scalar loses all visible structures and reaches a well-mixed state (orange, Fig.~\ref{fig:examples_mixing}a). Similarly, for low Damk{\"o}hler numbers (${\rm Da}=0.1$, first row in Fig.~\ref{fig:examples_mixing}b), the initial circle of the blue strain gets quickly stirred into the surroundings and loses all visible structure, leading to a well-mixed state with $f=0$ (red), despite the selective advantage of the blue strain. Weaker fluid transport (${\rm Da}=5$) eventually allows the selectively favored strain to win, as the initial circle maintains its interfacial integrity while being constantly folded and stirred. In contrast, advected model B dynamics generates phase separated domains that are broken up by turbulent mixing (Fig.~\ref{fig:examples_mixing}c), leading to arrested coarsening of drops at the so-called Hinze scale \cite{hinze_fundamentals_1955,perlekar_spinodal_2014}. Crucially, unlike the examples with conserved concentrations (Fig.~\ref{fig:examples_mixing}a,c), in the non-conserved system (Fig.~\ref{fig:examples_mixing}b), one of the strains always wins and achieves fixation at long times by filling the whole domain - the final states associated with Eqs.~\ref{eq:dcAdt}, \ref{eq:dcBdt} are thus an absorbing boundary condition at long times!

These examples demonstrate another key result - the Damk{\"o}hler number, and thus turbulent mixing, controls which strain outcompetes the other at fixed initial fractions and can override the selective advantage.
Below, we construct an analytical model to explain these results and test our predictions using numerical simulations.

\begin{figure}[t]
\centering
\includegraphics[width=\linewidth]{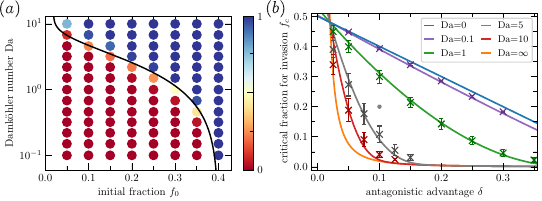}
\caption{
In (a), we show the fraction at fixation averaged over $N=50$ runs with different flow fields as a function of initial fractions $f_0$ and Damk{\"o}hler numbers with $\delta = 0.1$ and $\sigma=0.25$. The solid line is obtained from \eqref{eq:fcrit_ana}. Here, the factor $\alpha$ in the mixing time is obtained by fitting \eqref{eq:fcrit_ana} (solid lines) to the numerically measured $f_c$ in (b). Furthermore we show $f_c^\infty$ (\eqref{eq:fcrit_0}) and $f_c^0$ (\eqref{eq:fcrit_infty}) as function of $\delta$.
The grey dot denotes the parameter setting of Fig.~\ref{fig:examples_mixing}, predicting the transition of fixation between Da=1 and Da=5.
}
\label{fig:fixation_prob}
\end{figure}

\section*{Critical fraction for fixation under growth and mixing}
To understand how turbulent flows affect the criteria for fixation, we first consider two limiting cases. When turbulent mixing is much faster than biological reproduction (${\rm Da}\approx0$), the system rapidly becomes spatially homogeneous, i.e., the dynamics is effectively zero-dimensional. In this limit, setting $c_T\approx 1$ ($\sigma\ll 1$), we find that the dynamics of $f$ (\eqref{eq:dfdt}) has two stable fixed points at $f=0,1$ and an unstable fixed point
\begin{equation}
    f_c^0=\dfrac{2\sigma-\delta}{4\sigma}\;, \label{eq:fcrit_0}
\end{equation}
that sets the critical fraction required for fixation of strain A \cite{rouhani1987speciation,lavrentovich2019nucleation}.

In the opposite limit of vanishing flow (${\rm Da}=\infty$), we neglect advection in \eqref{eq:dfdt} and obtain the classical dynamics of model A (with $c_T \approx 1$). The spatial dynamics of invasion is now governed by a pushed genetic wave that connects domains of strain B ($f=0$) and strain A ($f=1$) with an interface width $w=2\sqrt{D/(\mu\sigma)}$ and speed $\vp = (\delta/2)\sqrt{D \mu/{\sigma}}$  \cite{murray2003mathematical,tanaka_spatial_2017}. 
As antagonism penalizes the contact between the two strains, the interface has an effective line tension $\gamma=(1/6)\sqrt{D\mu\sigma}$ (in 2D) \cite{rouhani1987speciation,lavrentovich2019nucleation} that leads to a nucleation barrier, and thus minimal domain size, required for a strain with selective advantage to colonize a less fit population. To see this, consider the dynamics of a circular domain of radius $R$ with a sharp interface ($w\ll R$) \cite{benzi_spatial_2022}, 
\begin{equation}  
    \dfrac{\dd R}{\dd t}=-\dfrac{D}{R}+\vp \;, \label{eq:dRdt}
\end{equation}
which (upon neglecting the periodic boundary conditions) defines the critical nucleation threshold $\Rcrit^\infty=(2/\delta)\sqrt{D \sigma/\mu}$ in the absence of mixing. In a system of size $L$, this yields a critical fraction (area fraction in the system: $f=\pi R^2/L^2$) for invasion to be
\begin{equation}
    \fcrit^\infty=\frac{4\pi  \sigma D }{\delta^2 L^2} \;, \label{eq:fcrit_infty}
\end{equation}
in the limit of ${\rm Da}=\infty$.

Motivated by the observations in Fig.~\ref{fig:examples_mixing}, we minimally combine the two limits for arbitrary ${\rm Da}$.
We assume that at early times, with respect to the eddy turnover time, fluid shear 
is unlikely to 
destroy spatial structures, and at late times, turbulent mixing effectively homogenizes the population. We neglect scale dependent fluctuations of the flow and quantify mixing by a single characteristic time scale
\begin{equation}
    \tmix = \alpha \tau_\text{eddy} \;, \label{eq:tmix}
\end{equation}
where the dimensionless parameter $\alpha$, which we expect to be of order unity, captures the effectiveness of turbulent mixing and can depend on biological parameters. Thus an initial circular domain of size $R_0$ is assumed to grow or shrink via a pushed wave (\eqref{eq:dRdt}) for $t<\tmix$, while for $t>\tmix$, the population is considered completely mixed with $\fcrit^0$ (Eq.~\ref{eq:fcrit_0}) now determining which strain reaches fixation. 

Direct integration of \eqref{eq:dRdt} yields
\begin{align}
    R(t,R_0)=\Rcrit^\infty 
    \Bigg[ 1 +  W_0 \left(\left( \frac{R_0}{\Rcrit^\infty}-1\right)
e^{\left(\frac{ R_0 + \vp t}{\Rcrit^\infty}-1\right)}  
\right) \Bigg], 
\label{eq:R_of_t}
\end{align}
as the domain radius at time $t$, where $W_0(x)$ is the principal branch of Lambert $W$-function. Note that for $R_0/\Rcrit^\infty <1$, the initial domain is below the nucleation threshold, and extinction occurs in a finite time $t_\text{ext}$ given by $t_\text{ext}=(\Rcrit^\infty/\vp )[-\ln(1-R_0/\Rcrit^\infty)- R_0/\Rcrit^\infty]$. Upon evaluating Eq.~\ref{eq:R_of_t} for $t=\tmix =\alpha \tau_\text{eddy}$, we can obtain the critical size threshold $R_\text{c}$ for the initial domain as an implicit relation
\begin{equation}
    \fcrit^0 = \frac{ \pi R(\tmix,R_\text{c})^2}{L^2}\;.
    \label{eq:fcrit_ana}
\end{equation}
To convert the critical domain size $R_c$ into a critical fraction, we use $\fcrit=\pi R_\text{c}^2/L^2$ and solve for $\fcrit$ to find
\begin{align}
    \fcrit = &\fcrit^\infty  \times \nonumber \\
    \Bigg[ 
     1 &+ W_0\left( \left( \sqrt{\frac{\fcrit^0}  {\fcrit^\infty}} -1 \right)e^{\left( \sqrt{\frac{\fcrit^0}  {\fcrit^\infty}} -1 - \frac{ \vp^2 \tmix}{ D}\right)} \right)
    \Bigg]^2\;. \label{eq:fcrit_ana}
\end{align}
Note that due the definition of the Lambert $W$-function, $W_0(\exp(\sqrt{\fcrit^0/\fcrit^\infty}-1)(\sqrt{\fcrit^0/\fcrit^\infty}-1)) = \sqrt{\fcrit^0/\fcrit^\infty}-1$ when $\tmix=0$, thus $\fcrit = \fcrit^0$ in this limit (${\rm Da}=0$). On the other hand, when  $\tmix=\infty$ (${\rm Da}=\infty$), $W_0(0) =0 $, thus $\fcrit = \fcrit^\infty$.


To test the validity and usefulness of this simplified theoretical model, we run numerical simulations of the continuum dynamics (Eqs.~\ref{eq:dcAdt},~\ref{eq:dcBdt}) for $\delta=0.1$, $\sigma=0.25$ and construct a phase diagram showing the final fraction at fixation for varying ${\rm Da}$ and initial fractions $f_0$ of strain A inoculated in the system (Fig.~\ref{fig:fixation_prob}a). For low Damk{\"o}hler numbers (${\rm Da}\to 0$), the invading species achieves fixation only when the initial fraction $f_0\geq f_c^0\approx 0.4$, but as Da increases, the critical fraction for invasion decreases sharply until it reaches $f_c\approx f_c^\infty$.
By averaging over $N=50$ runs, we compute the probability of fixation as a sigmoidal function of the selective advantage and identify the critical fraction $f_c$ from the inflection point (see Sec.~2 in the S.I.~and Fig.~S1 for details). Fig.~\ref{fig:fixation_prob}(b) shows how the critical fraction needed for fixation of strain A varies with selective advantage $\delta$, for different values of Da.
As expected, $f_c$ decreases as $\delta$ increases. 

Remarkably, our analytical prediction of the critical fraction \eqref{eq:fcrit_ana} agrees quantitatively with the numerical results (solid lines in Fig.~\ref{fig:fixation_prob}), with just a single parameter $\alpha = 3.4\pm0.2$ that we fit for across all Damk{\"o}hler numbers. That $\alpha>1$ highlights the fact that antagonistic mixtures take significantly longer than freely diffusing scalars to get mixed by turbulence; the latter typically mix within one large eddy turnover time \cite{frisch1995turbulence}.

Clearly, the simple analytical argument of initial growth followed by instantaneous mixing is a huge simplification of the real dynamics in these mixtures. Nevertheless, the overall shape of \eqref{eq:fcrit_ana} predicts the critical fraction for invasion effectively. Furthermore, we can predict the change in fixation probability between the two strains in Fig.~\ref{fig:fixation_prob}(a) quantitatively (black line), given the value of $\alpha$.
Although the analysis uses an initial circular geometry, our predictions and results are robust to different initial conditions (see S.I.~for details), highlighting the general validity of our theoretical framework.




\section*{Fixation time and mixing time}
Motivated by the success of our analytical model, we now use it to predict the dynamics of the system. In addition to the hydrodynamically important mixing time $\tmix$, a dynamical observable of considerable biological interest is the average fixation time $\tfix$, i.e., the time taken for either strain to reach fixation.
As we have a continuum model, it is necessary to define a criterion of fixation or conversely, extinction. We set the threshold for extinction to be when the global average of the dying strain is less than $\epsilon = 10^{-7}$, but our results are insensitive to this cutoff.

\begin{figure*}[t]
\centering
\includegraphics[width=0.66\linewidth]{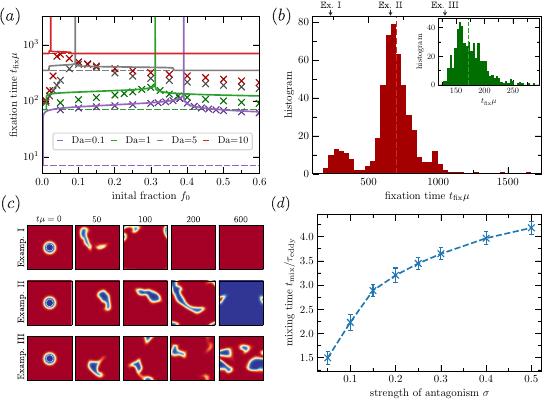}
\caption{
Fixation time as a function of the initial fraction $f_0$ for different Damk{\"o}hler numbers in units of the biological time scale $1/\mu$ (a). Points are obtained via the sample mean of $N=250$ runs with their corresponding standard error. Solid lines follow \eqref{eq:tfix_ana}. Dashed lines indicate corresponding mixing times $\tmix$.
In (b), we show the histogram of fixation times for $N=500$ runs for ${\rm Da = 10}$, $f_0 = 0.05$, and ${\rm Da = 1}$, $f_0 = 0.3$ in the inset.
In (c), we show spatial snapshots of the fraction under different flow fields for ${\rm Da = 10}$, $f_0 = 0.05$.
Measured mixing times as a function of $\sigma$ are shown in (f).
}
\label{fig:fix_times}
\end{figure*}

We measure the fixation time in numerical simulations of Eqs.~(\ref{eq:dcAdt},\ref{eq:dcBdt}) for different Damk{\"o}hler numbers. Fig.~\ref{fig:fix_times}(a) shows the average fixation time $\tfix$ (over $N=250$ runs) in units of the biological time scale $1/\mu$, for $\delta=0.1$, $\sigma=0.25$ and varying flow fields (Da)  and initial fractions ($f_0$).
In all cases, smaller Damk{\"o}hler numbers (i.e., stronger mixing) lead to shorter times for one species to reach fixation in the biological time scale. Close to the critical fraction ($f_0\approx \fcrit^0$), the mean fixation time reaches a sharp maximum and displays strong sample-to-sample fluctuations for low initial fractions and larger Damk{\"o}hler numbers (${\rm Da}=5,10$).

To understand these results, we adapt our previous theoretical arguments and estimate the fixation time in two steps (with $c_T\approx 1$). After one mixing time ($t=\tmix$), an initial fraction $f_0$ corresponding to a domain of size $R_0$, evolves to $f^\infty(\tmix,f_0)=\pi R(\tmix,R_0)^2/L^2$. The population is then assumed to be well-mixed for $t>\tmix$. Upon integrating \eqref{eq:dfdt} without spatial gradients, we obtain
\begin{align}
    &t^0(f,f^{\infty}) = \frac{2}{\mu} \times \nonumber  \\
    &\frac{(\delta+2\sigma)\ln\left(\frac{1-f}{1-f^\infty} \right) - 4\delta \ln\left(\frac{\delta+2\sigma(f-1)}{\delta+2\sigma(f^\infty-1)} \right) - (\delta-2\sigma)\ln\left(\frac{f}{f^\infty} \right)}{\delta^2-4 \sigma^2},
\end{align}
for the time it takes a starting fraction $f^\infty$ to reach $f$.
We assume $|\delta|< 2 \sigma$, thus avoiding the spinodal lines for this phase separation problem at $\alpha=0$ and $\beta =0$.
By taking into account which strain goes to fixation depending on the critical fraction, we combine these results to obtain the fixation time as
\begin{equation}
    \tfix = 
    \begin{cases}
        \tmix + t^0(\epsilon, f^\infty(\tmix, f_0)) \;, \text{ if } f^\infty(\tmix, f_0) < f_c^\infty \\
        \tmix + t^0(1-\epsilon, f^\infty(\tmix, f_0)) \;,  \text{ otherwise}
\end{cases} . 
\label{eq:tfix_ana}
\end{equation}

We plot the predicted fixation time (Eq.~\ref{eq:tfix_ana}) in Fig.~\ref{fig:fix_times}(a) (solid curves) and find excellent agreement for ${\rm Da\leq 1}$ with no free parameters ($\alpha$ is obtained from the fits for the critical fraction, see Fig.~\ref{fig:fixation_prob}). 
Interestingly, the dynamics for ${\rm Da}>1$ and ${\rm Da}\leq 1$ are qualitatively different. In the latter case, mixing is sufficiently strong for antagonism to delay fixation. As a result, the measured $\tfix>\tmix$ ($\tmix$ is indicated by the dashed horizontal lines according to the Damk{\"o}hler numbers in Fig.~\ref{fig:fix_times}(a), and our analytical prediction (Eq.~\ref{eq:tfix_ana}) gets better over a wide range of initial fractions $f_0$, as the Damk{\"o}hler number decreases.
Furthermore, our numerical results also show the highest fixation times close to the critical fraction of fixation, as predicted by the logarithmic divergence in \eqref{eq:tfix_ana}. These times result because for low Damk{\"o}hler numbers $f$ is well-mixed and close to the unstable fixed point $f_c^0$.
In contrast, for larger Damk{\"o}hler numbers (${\rm Da}>1$), fixation occurs quickly with the measured $\tfix < \tmix $, which our analytical result (\eqref{eq:tfix_ana}) fails to capture. In this regime, fluctuations in the flow become important for the dynamics, especially near the critical fraction, and our approximation of a single mixing timescale breaks down. 

To better understand these fluctuations and their impact on the fixation time, we show a histogram ($N=500$ different runs) of fixation times for ${\rm Da = 10}$ ($f_0 = 0.05$) in the main panel of Fig.~\ref{fig:fix_times}(b). Interestingly, in this case, we find a bimodal distribution with a smaller peak for much lower fixation times then the second peak, which is spread around the prediction from Eq.~\ref{eq:tfix_ana} (dashed vertical line). Furthermore, there exist some rare events of extreme long fixation times. Both of these features are absent in the histogram of fixation times for ${\rm Da = 1}$ ($f_0 = 0.3$), shown in the inset of Fig.~\ref{fig:fix_times}b. Here, we find a single peak around the predicted time (via Eq.~\ref{eq:tfix_ana}, indicated by the dashed vertical line). 
Remarkably, for ${\rm Da}>1$, our analytical approach is able to capture the {\it most probable} fixation time, rather than the average, as shown in the main panel of Fig.~\ref{fig:fix_times}(b).

We show snapshots of the dynamics for three different flow realizations in Fig.~\ref{fig:fix_times}(c) at $\rm Da=10$, and indicate their fixation time on top of Fig.~\ref{fig:fix_times}(b)s.
In some cases, the advected domains get stretched and quickly split below their critical nucleation radius (indicated by the white circle), allowing strain B to immediately fix (see Fig.~\ref{fig:fix_times}(b) - Examp.~I; first row). In other cases, 
intermittent fluctuations in the flow remain weak enough to only deform the domains a little; providing strain A sufficient time to grow and invade the system (see Fig.~\ref{fig:fix_times}(b) - Examp.~II; second row).
In still other cases, consecutive splitting, and partial extinction events can maintain drops of the invading strain close to the critical size, causing a dramatic increase in the fixation time (see Fig.~\ref{fig:fix_times}(b) - Examp.~III; third row).
All these different scenarios lead to the broad range of fixation times for high Damk{\"o}hler numbers.
In some cases, with relatively low probability, the biological reproduction is fast enough to keep the integrity of domain interfaces, and thereby preventing mixing. For such domains, fixation can now happen on a time scale below the mixing time, set by a pushed-wave dynamics of these interfaces, and the homogenizing effect of mixing gets lost. Therefore, in these cases,  the specific flow fields matter much more for the fate of a domain.

Finally, we address the dependence of the mixing time ($\tmix$) on biological antagonism ($\sigma$). Intriguingly, we find that the measured $\tmix$ (obtained by fitting for the critical fraction as in   Fig.~\ref{fig:fixation_prob}(b)) increases with antagonism (Fig.~\ref{fig:fix_times}(d)). An order of magnitude increase in $\sigma$ (from 0.05 to 0.5) leads to a nearly 3-fold increase in the mixing time, quantifying our intuition that antagonism makes mixing by stirring harder.  

\section*{Concentration spectra under turbulent mixing}
So far, we focused mainly on the global dynamics of the system. We now quantify spatial fluctuations of the competing populations and draw comparisons with classical studies of turbulent passive scalars or phase-separating mixtures \cite{ruiz1981turbulence,shraiman2000scalar,berti2005turbulence,perlekar_spinodal_2014}.

We measure the shell-averaged concentration spectrum normalized by the corresponding structure function $S(k_n) = (\sum_{k_n}' | \hat{f}(\mathbf{k},t)|^2)/(\sum_{k_n}'1)$, where $\hat{f}(\mathbf{k},t)$ is the Fourier transform of the fraction $f(\mathbf{x},t)$ and $\sum_{k_n}'$ denotes the sum over all modes within the $n$'th shell, i.e., a mode with modulus $k=|\mathbf{k}|$ lies in the $n$'th shell when  $k \in [k_0 b^{n-1/2}, k_0 b^{n+1/2}]$, where $b$ is a discretization parameter and $k_0 = 1/L$ (smallest mode). We choose $b=1.14$, such that we obtain roughly five shells per octave. We similarly compute the conventional energy spectrum $E(k)$ of fluid turbulence \cite{frisch1995turbulence}.

\begin{figure*}[t]
\centering
\includegraphics[width=0.66\linewidth]{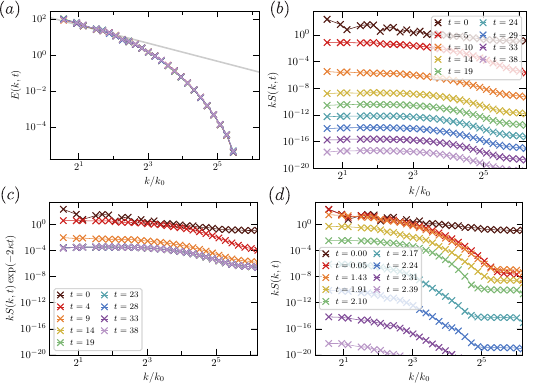}
\caption{Shell averaged spectra of the kinetic energy (a) at the statistical steady state and the fraction $f$ of genetic mixtures with Da=0.1 (b) and Da=10 (d) at selected time points. In (c), we show the exponentially rescaled spectrum of (b).
All time scales are given in the time scale of reproduction $1/\mu$.
}
\label{fig:spectra}
\end{figure*}
Fig.~\ref{fig:spectra}(a) shows the stationary state of the energy spectrum for our 2D incompressible, turbulent flow field, along with the Kolmogorov scaling $E(k) \propto k^{-5/3}$, indicated as a straight line. 

In Figs.~\ref{fig:spectra}(b,d), we show the shell-averaged concentration spectrum of the transient dynamics for genetic mixtures with Da=0.1 and 10 at different time points in the evolution of a circular disc inoculation, averaged over $N=100$ runs with different flow fields.
As the final state of an antagonistic mixture is an absorbing state with $f=0$ or $f=1$, all spatial fluctuations eventually vanish, causing the spectra to converge to zero in the long time limit. However, the transient dynamics depends strongly on the Damk{\"o}hler number. Larger Damk{\"o}hler numbers cause short wavelength fluctuations to get more strongly damped. 
Furthermore, for large Damk{\"o}hler numbers, after an initial continuous reduction of the amplitude, the whole spectrum disappears quite fast around the fixation time. This abrupt disappearance reflects the fact that for large Da, fixation is reached largely by the growth and shrinkage of domains. 
For lower Damk{\"o}hler numbers, however, the domains get mixed, and the system is more homogeneous already in the transient regime.
This homogenization is seen more clearly when comparing the concentration spectrum of the genetic mixture to that of a passive scalar.

To highlight this relation, in Fig.~\ref{fig:spectra}(c), we exponentially rescale in time the spectrum of the ${\rm Da}=0.1$ genetic mixture, which shows good collapse at late times, similar to that of the passive scalar. This collapse can be understood through a simple linearized analysis about one of the final absorbing states, valid for low Damk{\"o}hler numbers and in the long time limit. Upon writing $\psi(x,t) = f(x,t) - f^*$, we obtain $(\partial_t+\mathbf{v}\cdot\nabla)\psi =D \nabla^2 \psi - \kappa \psi$, where $\kappa = \mu ( \sigma \pm \delta/2 )$ for $f^*=0$ or $f^*=1$. The transformation $\Psi(x,t) = \exp(\kappa t)\psi(x,t)$ converts this linear dynamics to that of an advected passive scalar, namely $(\partial_t+\mathbf{v}\cdot\nabla)\Psi =D \nabla^2 \Psi $, thus demonstrating the suggested equivalence.
\section*{Discussion}
The growth of killer microbe strains in a turbulent environment warrants an unusual combination of conflicting tendencies - biological antagonism that actively tries to segregate different strains and turbulent stirring that forces intermingling of the antagonistic populations. Our analysis highlights the role of nucleation thresholds in fixation and demonstrates how turbulent mixing strongly dictates which strain survives (or goes extinct), when it does so, and how its dynamics evolves spatially. For phytoplankton in the ocean with typical growth rates $\mu\sim 0.5-1~{\rm day}^{-1})$ and eddy turnover times $\tau_{\rm eddy}\sim 5$~minutes to $\gtrsim50~$days (on a range of spatial scales, $\sim 0.1-100~$km) \cite{martin2003phytoplankton,mahadevan2016impact}, a wide range of ${\rm Da}\sim 10^{-3}-10^4$ is possible, potentially allowing an arena to test our predictions. 

A key ingredient in our work is the presence of nonreciprocal ($\alpha\neq\beta$) and inhibitory ($\alpha,\beta<0$) interactions in the growth rates of the organisms, making the system a kind of active matter \cite{hallatschek2023proliferating}. Nonreciprocal interactions are known to generate spatial patterns in microbial mixtures, e.g., motile bacteria and phages \cite{martinez2023pattern} and cause chaotic evolution with sustained diversity in multicomponent communities \cite{pearce2020stabilization}. Whether such mechanisms of sustaining diversity survive in the presence of fluid turbulence and the absorbing boundary conditions enforced by fixation remains a challenging problem for the future, as do understanding the effect of compressible turbulent velocity fields and generalizing our results to three dimensions.

In the limit of reciprocal interactions ($\alpha=\beta$, so $\delta=0$), our model (Eqs.~\ref{eq:dcAdt},~\ref{eq:dcBdt}) reduces to similar autocatalytic models describing the synthesis of chiral biomolecules and the origins of biological chirality \cite{frank1953spontaneous,jafarpour2015earlylife,brandenburg2004long}. That turbulent transport and fluid dynamical mixing effects can speed up the attainment of a fully homochiral state in chemical systems \cite{kondepudi1990chiral} may play an important role in dictating the handedness of prebiotic mixtures on distant planets \cite{brandenburg2004long}. Our results show that the effects of a small chiral bias ($\delta\neq 0$) due to stray magnetic fields, fundamental parity violations, etc., \cite{saito2013colloquium} can be strongly influenced by turbulent flow, suggesting that the physical environment of water may be as important as its chemical nature in the evolution of early life.


\matmethods{
Extended data on the details of numerically generating the flow field. Further details on the model, theoretical calculations and analysis are provided in the Supplementary Information.
\subsection*{Data and code availability}
Code used for the numerical simulations will be made available on a public repository upon publication. All other data needed to reproduce the results in this paper are provided in the Methods and Supplementary Information.
\subsection*{Generation of the flow fields}
We use a pseudo-spectral method for solving the Navier-Stokes equation in a three-dimensional lattice with periodic boundary conditions with random forcing at long scales. In all our simulations, we have chosen the same resolution ($N=128$ grid points for a system size with $L=2 \pi$) as for the numerical solver for the advected reaction-diffusion equations later.
However, in this work, we focus on the flow of a two-dimensional (2D) slice originating from the three-dimensional (3D) flow field. This choice is motivated by the fact that many oceanographic micro-organisms control their buoyancy and live in a height-selected layer of the ocean; see \cite{martin2003phytoplankton,benzi_spatial_2022}.
By taking the two-dimensional slice, we are generally left with a compressible flow. In this work, however, we restrict ourselves to the easier situation of an incompressible flow for simplicity, leaving the interesting effects of compressibility on the mixing of antagonistic populations for future work.
Following the Ref.~\cite{perlekar_2013_comp}, we obtain the 2D incompressible part of the flow field $\mathbf{v}$ by projecting its Fourier transform $\tilde{\mathbf{v}}$ via
\begin{equation}
    \mathbf{v}_\text{incom.} = \mathcal{F}^{-1}
    \left( 
    \tilde{\mathbf{v}} - \frac{\mathbf{k} (\mathbf{k} \cdot \tilde{\mathbf{v}})}{|\mathbf{k}|^2}
    \right), 
\end{equation}
where $\mathcal{F}^{-1}(\mathbf{x})$ denotes the inverse Fourier transform of $\mathbf{x}$, and $\mathbf{k}$ is the frequency variable. Note that with this projection, we insure $\nabla \cdot \mathbf{v}_\text{incom.} = 0$, because $\mathbf{k} \cdot \tilde{\mathbf{v}}_\text{incom.}  = \mathbf{k} \cdot \tilde{\mathbf{v}} - \mathbf{k} \cdot \tilde{\mathbf{v}}$, where $\tilde{\mathbf{v}}_\text{incom.}$ is the Fourier transform of the incompressible flow field $\mathbf{v}_\text{incom.}$. The latter is just called $\mathbf{v}$ in the main text.
}

\showmatmethods{} 

\acknow{
All authors acknowledge Martijn Dorrestijn for his contribution in the early stages of this project.
J.B.~and D.R.N.~acknowledge support from the Harvard Materials Research Science and Engineering Center through National Science Foundation grant DNR-2011754.
J.B.~thanks the German Research Foundation for financial support through the DFG Project BA 8210/1-1. S.S.~acknowledges illuminating discussions during the ``Anti-Diffusion in Multiphase and Active Flows'' hosted by the Isaac Newton Institute for Mathematical Sciences, Cambridge, supported by the EPSRC grant EP/R014604/1.
}







\showacknow{} 



\end{document}